\documentclass[superscriptaddress,twocolumn,amssymb,nofootinbib,10pt]{revtex4-2}
\usepackage[normalem]{ulem}
\usepackage[utf8]{inputenc}
\usepackage{comment}
\usepackage{graphicx}
\usepackage{graphics}
\usepackage{amsmath, amssymb}
\usepackage{graphics,bm}
\usepackage{graphicx}
\usepackage{bbold}
\usepackage{slashed}
\usepackage{feynmf}
\usepackage{wrapfig}
\usepackage{hyperref}
\usepackage{cancel}
\usepackage[usenames]{color}  
\usepackage{lipsum}
\usepackage[compat=1.1.0]{tikz-feynman}
\usepackage{tikz}
\usetikzlibrary{arrows,shapes}
\usetikzlibrary{trees}
\usetikzlibrary{matrix,arrows} 			
\usetikzlibrary{positioning}				
\usetikzlibrary{calc,through}				
\usepackage{pgffor}							

\newcommand{\be}{\begin{equation}}
\newcommand{\ee}{\end{equation}}
\newcommand{\beq}{\begin{equation}}
\newcommand{\eeq}{\end{equation}}

\newcommand{\bqa}{\begin{eqnarray}}
\newcommand{\eqa}{\end{eqnarray}}

\newcommand{\ms}{\overline{\text{\tiny MS}}}

\def\square{\vcenter{\vbox{\hrule height.4pt
          \hbox{\vrule width.4pt height4pt
          \kern4pt\vrule width.3pt}\hrule height.4pt}}}

\usepackage[normalem]{ulem}
\newcommand{\nn}{\nonumber}

\begin{document}

\title{Pion condensation versus 2SC, speed of sound, and charge neutrality effects in the quark-meson diquark model}
\author{Jens O. Andersen}
\email{jens.andersen@ntnu.no}
\affiliation{Department of Physics, Faculty of Natural Sciences,NTNU, 
Norwegian University of Science and Technology, H{\o}gskoleringen 5,
N-7491 Trondheim, Norway}

\author{Mathias P. N{\o}dtvedt}
\email{mathias.p.nodtvedt@ntnu.no}

\affiliation{Department of Physics, Faculty of Natural Sciences,NTNU, 
Norwegian University of Science and Technology, H{\o}gskoleringen 5,
N-7491 Trondheim, Norway}

\date{\today}

\begin{abstract}
We employ the two-flavor quark-meson diquark model as a low-energy 
model for QCD at non-zero quark and isospin chemical potentials
$\mu$ and $\mu_I$, and at zero temperature. We map out the phase diagram in the $\mu$-$\mu_I$ plane, which has four phases: a vacuum phase, a 
phase with condensed charged pions/Cooper pairs of $u$ and $\bar{d}$
quarks, a normal quark matter phase, and a color superconducting phase (2SC phase). 
The global symmetry breaking $SU(3)_c\rightarrow SU(2)_c$
in the 2SC phase gives rise to a number of Nambu-Goldstone bosons. We classify them and briefly discuss their properties.
We calculate the speed of sound $c_s$ in the two special cases, 
finite $\mu_I$ and $\mu=0$, and finite $\mu$ and $\mu_I=0$.
In both cases, the speed of sound exhibits a maximum and approaches the conformal limit from above as the density increases. For non-zero isospin $\mu_I$, this behavior is in agreement with the speed of sound obtained from lattice simulations. In the 2SC phase, the behavior is qualitatively
the same if we impose local charge neutrality.
Finally, we discuss the possibility of having a mixed phase of negatively charged normal quark matter and positively charged 2SC matter with global color charge neutrality imposed on the latter.

\end{abstract}

\maketitle
\section{Introduction}
QCD in extreme conditions has been of interest for decades due to its
relevance to the early universe, heavy-ion collisions, and 
compact stars~\cite{raja,alford,fukurev}.
The QCD phase diagram, which conventionally is drawn in the $\mu$--$T$
plane, turns out to be very rich, with a number of phases.
It becomes even more complicated if we allow an independent chemical potential for each quark flavor $f$ instead of a common baryon chemical potential. For two quark flavors, we can then use $\mu_u$ and $\mu_d$
or equivalently use the isospin chemical potential $\mu_I=\mu_u-\mu_d$
and the baryon chemical potential $\mu_B=3\mu$. For three flavors, we add a $\mu_S$-axis, where $S$ is strangeness, and $\mu_S={1\over2}(\mu_u+\mu_d-2\mu_s)$.
Consider the phase diagram in the $\mu_B$--$\mu_I$ plane.
Moving along the $\mu_I$-axis, QCD is in the vacuum phase until the isospin chemical potential reaches the critical value $\mu_I^c=m_{\pi}$ where one expects the onset of charge pion condensation~\cite{sonstep}. For sufficiently small values of $\mu_I$, the pion-condensed phase can be described in a model-independent way using chiral perturbation theory~\cite{chipt,chipt2}. For very large values of $\mu_I$, quarks and gluons are the relevant degrees of freedom, and if the density is sufficiently high, one can use perturbative QCD (pQCD). In pQCD, one-gluon exchange gives rise to an attractive quark-quark interaction channel. According to BCS theory for superconductivity, any attractive interaction renders the Fermi sphere unstable, and Cooper pairs are formed. 
Since $\mu_u=-\mu_d$, these pairs consist of $u$ quarks and $\bar{d}$
antiquarks or vice versa depending on the sign of $\mu_I$.
Thus, QCD at very large isospin chemical potential is a color superconductor. The diquark condensate or the gap has the same quantum numbers as those of the pion condensate that exists for low values of $\mu_I$, so one expects a BEC-BCS crossover transition as one increases $\mu_I$~\cite{sonstep}. A nice feature of QCD at non-zero $\mu_I$ but $\mu_B=0$ is the absence of the sign problem and the possibility to perform lattice simulations. It is then possible to compare lattice QCD~\cite{lattice,abb23,abb24} with model-independent predictions for small $\mu_I$ using chiral perturbation theory~\cite{sonstep,qing} and perturbative QCD for large $\mu_I$.

The situation is similar at large baryon chemical potentials, where perturbative QCD can be applied. Again, there is an attractive channel that involves one-gluon exchange. For sufficiently large values of $\mu_B$, all three quark flavors can be taken massless, and they participate in the pairing on an equal footing as do the quarks of different colors. The resulting ground state is referred to as the color-flavor-locked (CFL) phase of QCD, whose existence at asymptotically
large  baryon chemical potential is one of the few rigorous results
in cold and dense QCD. Moving towards smaller values of $\mu_B$, one can no longer ignore the mass of the $s$-quark. The finite value of $m_s$ places stress on pairs that contain an $s$-quark. Eventually, it is energetically advantageous to break up such pairs, and one is left with
a pairing pattern that involves the $u$ and $d$ quarks and two of the three colors. This phase is referred to as the 2SC phase.~\footnote{Before
the breakup of Cooper pairs, the stress on them may be relieved
by kaon condensation. The system is still in the CFL phase, but
the condensation corresponds to a rotation of the order parameter~\cite{bedaque2}.}
Note, however, that the existence of the 2SC phase depends on the details of the models employed. For example, in the NJL model, there is a
2SC phase between the phase of normal quark matter and the CFL phase,
if the couplings are small. For larger values of the couplings, the
transition from the CFL phase directly to normal quark 
matter~\cite{rajagopal1,rusterneutral,abuki}.

The interest in QCD at finite quark chemical potentials and low temperature is mainly due to its applications to compact stars. Knowing the equation of state allows us to calculate macroscopic quantities such as masses and radii of individual neutron stars as well as tidal deformabilities in binary systems. The recent observations of compact stars of more than two solar masses put constraints on the equation of state and one of the main questions is whether the most massive stars have a quark core surrounded by ordinary nuclear matter.
The quest for a phase transition in dense QCD has been hampered by the
the inapplicability of lattice QCD at finite $\mu_B$ due to the sign problem. The sign problem forces one to use other methods.
At very large values of the baryon chemical potential, perturbative QCD (pQCD) is applicable due to asymptotic freedom. In recent years, perturbative calculations have been pushed to order $\alpha_3^3\log\alpha_s$, where
$\alpha_s$ is the strong coupling~\cite{dense1,dense4}.
In Ref.~\cite{loic}, renormalization group methods have been used to improve the convergence of perturbative series.
For lower values of $\mu$, where pQCD cannot be applied, one uses 
low-energy effective models that in some aspects resemble QCD.
The Nambu-Jona-Lasinio model is an example of a low-energy effective
model that incorporates chiral symmetry breaking in the vacuum, but
has no dynamical gauge fields. The local $SU(3)_c$ gauge symmetry is
then replaced by the corresponding global symmetry.

The present work is a continuation of Ref.~\cite{us}. In that paper, we formulated the two-flavor quark-meson diquark (QMD) model as a renormalizable low-energy effective model for two-flavor QCD, with meson, quark, and diquark degrees of freedom. 
In this paper, we continue our study of the model and its properties,
focusing on the phase diagram in the $\mu_I$--$\mu_B$ plane, the speed of
sound, and the effects of imposing neutrality for electric and color charge.

\section{Two-flavor QMD model}

\subsection{Lagrangian}
The Minkowski Lagrangian for the two-flavor quark-meson diquark model is

\begin{widetext}

\begin{eqnarray}
\nonumber
{\cal L}&=&
{1\over2}(\partial_{\mu}\sigma)(\partial^{\mu}\sigma)
+{1\over2}(\partial_{\mu}\pi_0)(\partial^{\mu}\pi_0)
+\left(\partial_{\mu}+i\mu_I\delta_{\mu0}\right)\pi^+
\left(\partial^{\mu}-i\mu_I\delta^{\mu0}\right)\pi^-
-{1\over2}m^2(\sigma^2+\vec{\pi}^2)-{\lambda\over24}(\sigma^2+\vec{\pi}^2)^2
\\ \nonumber&&+h\sigma
+\left(\partial_{\mu}+2i{\mu}_a\delta_{\mu0}\right)
\Delta^{\dagger}_a\left(\partial^{\mu}-2i{\mu}_a\delta^{\mu0}\right)\Delta_a
-m_{\Delta}^2\Delta^{\dagger}_a\Delta_a
-{\lambda_3\over12}(\sigma^2+\vec{\pi}^2)\Delta^{\dagger}_a\Delta_a
-{\lambda_{\Delta}\over6}\left(\Delta^{\dagger}_a\Delta_a\right)^2
\\ &&
+\Bar{\psi}(i\slashed{\partial}+\gamma^0\hat{\mu})\psi-g\bar{\psi}[\sigma+i\gamma^5\vec{\pi}\cdot\vec{\tau}]\psi
+{1\over2}g_{\Delta}\Bar{\psi}^C_b\Delta_a\gamma^5\tau_2
\epsilon_{abc}\psi_c
+{1\over2}g_{\Delta}\Bar{\psi}_b\Delta_a^{\dagger}\gamma^5\tau_2\epsilon_{abc}
\psi_c^C\;, 
\label{lagrangian}
\end{eqnarray}
\end{widetext}
where $\psi$ is a flavor doublet and a color triplet in the fundamental representation
\begin{eqnarray}
\psi&=&
\left(\begin{array}{c}
\psi_{ur}\\
\psi_{dr}\\
\psi_{ug}\\
\psi_{dg}\\
\psi_{ub}\\
\psi_{db}\\
\end{array}\right)\; .
\end{eqnarray}
Furthermore, $\psi_c$ is the flavor doublet with color $c$
\begin{eqnarray}
\psi_c&=&
\left(\begin{array}{c}
\psi_{uc}\\
\psi_{dc}\\
\end{array}\right)\; .
\end{eqnarray}
The field $\Delta_a$ is a composite diquark field, $\Delta_a\sim\bar{\psi}_b\tau_2\epsilon_{abc}\gamma^5\psi_c^C$, where \(\psi^C = C\bar{\psi}^T\) and  $C=i\gamma^2\gamma^0$ is the charge conjugation operator, which
can be written as
\begin{eqnarray}
\Delta&=&
\left(\begin{array}{c}
\Delta_1\\
\Delta_2\\
\Delta_3\\
\end{array}\right)\;.
\end{eqnarray}
The diquark field $\Delta$ transforms as a singlet under $SU(2)_L\times SU(2)_R$ and as an antitriplet under $SU(3)_c$. Furthermore,
$\hat{\mu}$ is the quark chemical potential matrix, $\mu_I$ is the isospin chemical potential for electrically charged pions, and $\mu_a$ is the chemical potential for diquark $a$.
The chemical potentials $\hat{\mu}$, $\mu_a$, and $\mu_I$
are not independent, but can be expressed in terms of 
$\mu_B$, $\mu_e$, and $\mu_8$ which are the chemical potentials for baryon
number, electric charge, and color charge $Q_8$ (see subsection~\ref{neu}). 
The symmetries of the Lagrangian in Eq.~\eqref{lagrangian} were discussed in some detail in Ref.~\cite{us}, here we 
simply state them. In the chiral limit, the global symmetries are $SU(2)_L\times SU(2)_R\times U(1)_B\times SU(3)_c$, while at the physical point, the global symmetry is $SU(2)_V\times SU(3)_c$.
The diquark condensate $\Delta_0=\langle\Delta_3\rangle$ in the 2SC phase
breaks the $SU(3)_c$ to $SU(2)_c$. The $U(1)_B$ symmetry 
in the vacuum, which is generated by $B={1\over3}{\rm diag}(1,1,1)$
is modified in the medium, $\tilde{B}=B-{2\over\sqrt{3}}T_8={\rm diag}_c(0,0,1)$, where $T_a$ are the generators of  $SU(3)$. 
In the 2SC phase, only the blue quasiparticles carry a non-zero 
$\tilde{B}$ charge~\cite{igor}. Similarly, there is an unbroken modified $U(1)$ symmetry generated by $Q^{\prime}=Q-{1\over\sqrt{3}}T_8$, where $Q={\rm diag}({2\over3},-{1\over3})$ is the generator of the electromagnetic $U(1)$ symmetry.
In the QMD model, this symmetry is global, whereas in QCD it is local.
The gauge boson associated with this symmetry is then the in-medium photon.
In the condensed phase of the pion, the condensate $\rho_0$ breaks
the symmetry $U(1)_{I_3}$ associated with the third component of the
isospin, that is, the subgroup of $SU(2)_V$. The $U(1)$ electromagnetic symmetry is also broken since ${I_3}=Q+{1\over6}\mathbb{1}$. 

The Lagrangian has a number of masses and couplings. They can be grouped whether they belong to the quark-meson sector or the diquark-meson sector. The former include $m^2$, $\lambda$, $g$, and $h$, while the latter includes $m_{\Delta}^2$, $\lambda_3$, $\lambda_{\Delta}$, and $g_{\Delta}$.
The masses and couplings are all running parameters that depend on the renormalization scale $\Lambda$ and each of them satisfies a renormalization group equation.
These equations and their solutions i.e. the running parameters were derived in Ref.~\cite{us}. The running masses and couplings can be used to obtain an RG-improved effective potential.

\subsection{Chemical potentials and neutrality conditions}
\label{neu}
According to Noether's theorem, there is one conserved charge for each continuous symmetry of the system. In principle, one can introduce an independent chemical potential $\mu_i$ for each conserved charge $Q_i$.
The corresponding number density is denoted by $n_{Q_i}$.
However, this can only be done if the charges commute. In other words, we can introduce a maximum of $N$ independent chemical potentials corresponding to the maximum number $N$ of commuting charges of the
system. In the present case, these are the baryon chemical potential $\mu_B$, the chemical potential for electric charge $\mu_e$, and the
chemical potentials for the color charges $Q_3$ and $Q_8$, 
$\mu_3$ and $\mu_8$. The pairing of quarks in the 2SC phase gives rise to a non-zero color charge $Q_8$ in the ground state, while $Q_3=0$. 
Therefore, it is sufficient to introduce $\mu_8$.
This is in contrast to the CFL phase of three-flavor QCD, where the ground state completely breaks the $SU(3)_c$ symmetry. For this reason, both $\mu_3$ and $\mu_8$ must be introduced in order to impose
color charge neutrality.

The quark chemical potential matrix can be expressed in terms of $\mu_B$, $\mu_e$, and $\mu_8$ as  
\begin{eqnarray}
\mu_{ij,\alpha\beta}&=&(\mbox{$1\over3$}\mu_B\delta_{ij}-\mu_eQ_{ij})\delta_{\alpha\beta}
+\delta_{ij}
{2\over\sqrt{3}}\mu_8
(T_8)_{\alpha\beta} \;,
\label{chempot}
\end{eqnarray}
where $i,j$ are flavor indices and $\alpha,\beta$ are color indices, and
$Q_{ij}$ are the matrix elements  of the matrix
$Q={\rm diag}({2\over3},-{1\over3})$.
Alternatively, one could use the chemical potentials $\mu_B$ and $\mu_I$
instead with the replacement $(\mbox{$1\over3$}\mu_B\delta_{ij}-\mu_eQ_{ij})\rightarrow(\mbox{$1\over3$}\mu_B\delta_{ij}+{1\over2}\mu_I(\tau_3)_{ij}\delta_{\alpha\beta}$).
The expressions for the different quark chemical potentials follow from 
Eq.~(\ref{chempot}) as
\begin{eqnarray}
\mu_{ur}&=&\mu_{ug}=\mu-{2\over3}\mu_e+{1\over3}\mu_8\;,\\   
\mu_{dr}&=&\mu_{dg}=\mu+{1\over3}\mu_e+{1\over3}\mu_8\;,\\   
\mu_{ub}&=&\mu-{2\over3}\mu_e-{2\over3}\mu_8\;,\\
\mu_{db}&=&\mu+{1\over3}\mu_e-{2\over3}\mu_8\;,
\end{eqnarray}
where $\mu={1\over3}\mu_B$ is the  common quark chemical potential.
The number densities are given by 
\begin{eqnarray}
n_B&=&-{\partial \Omega\over\partial\mu_B}\;,  \\
n_e&=&-{\partial \Omega\over\partial\mu_e}\; ,  \\
n_{8}&=&-{\partial \Omega\over\partial\mu_{8}}\; ,  
\end{eqnarray}
where $\Omega$ is the thermodynamic potential.
For arbitrary values of $\mu_B$, $\mu_e$, and $\mu_8$, the electric charge density is non-zero, $n_e\neq0$. However, a macroscopic chunk of quark matter must be electrically neutral, otherwise a huge energy price has to be paid due to the Coulomb  interaction~\cite{igor}. We therefore impose local electric charge neutrality on the system. The same remarks apply to the color charge $Q_8$ and we therefore impose the two constraints 
\begin{eqnarray}
{\partial\Omega\over\partial\mu_e}={\partial\Omega\over\partial\mu_8}&=&0\;,
\label{neucond}
\end{eqnarray}
which finally leaves us with a single independent chemical 
potential, e. g. $\mu_B$. Electric charge neutrality is obtained
by adding an electron background, which is achieved by adding
the following term to the thermodynamic potential
\begin{eqnarray}
\Omega_e&=&-{4\mu_e^4\over3(4\pi)^2}\;.    
\end{eqnarray}

In nature, $U(1)_Q$ and $SU(3)_c$ are local symmetries, and the zeroth components of the gauge fields serve as chemical potentials. The values of these fields are dynamically driven to values that set these charges to zero~\cite{chempot,gerhold}. In the quark-meson diquark model, the above-mentioned symmetries are global, and the chemical potentials must be introduced by hand. In normal quark matter, it is straightforward
to derive that $\mu_8=0$ guarantees color charge neutrality.
However, in two-flavor QCD, normal quark matter is electrically charged
since the trace of the electric charge matrix is non-zero. This requires a non-zero $\mu_e$ and the presence of electrons.
In the 2SC phase, electric and charge color neutrality is ensured by
non-zero values of both $\mu_e$ and $\mu_8$, as we shall see below.

\subsection{Quark spectrum and thermodynamic potential in the 2SC phase}
We write the sigma $\sigma$
and diquark field $\Delta_3$ as a sum of their expectation
values $\phi_0$ and $\Delta_0$, and quantum fluctuating fields.
This is done by making the substitutions
\begin{eqnarray}
    \sigma &\rightarrow& \phi_0+{\sigma}\;,\\
    \Delta_3 &\rightarrow& \Delta_0+{\Delta}_3\;,
\end{eqnarray}
in the Lagrangian. Choosing $\Delta_0$ real, the resulting tree-level thermodynamic potential is
\begin{widetext}
\begin{eqnarray}
\Omega_0&=&{1\over2}m^2\phi_0^2+(m_{\Delta}^2-4\bar{\mu}^2)\Delta_0^2
+{\lambda\over24}\phi_0^4
+{\lambda_3\over12}\phi_0^2\Delta_0^2+{\lambda_{\Delta}\over6}\Delta_0^4-h\phi_0\;, 
\label{treepot0}
\end{eqnarray}
where $\bar{\mu}$ is defined in Eq.~(\ref{mubar}) below.
The bilinear quark terms in the Lagrangian are
\begin{eqnarray}
{\cal L}_{\rm bilinear}
&=&\bar{\psi}[i\slashed{\partial}-g\phi_0+\hat{\mu}\gamma^0]\psi
+\frac{1}{2}g_\Delta\bar{\psi}^C_b\Delta_0\gamma^5\tau_2\epsilon_{ab3}\psi_c
     +\frac{1}{2}
g_\Delta\bar{\psi}_b\Delta_0^\dagger\gamma^5\tau_2\epsilon_{ab3}\psi_c^C\;. 
\end{eqnarray}
Using the Nambu-Gorkov basis, \(\Psi = (\psi, \psi^C)\), we then find the following inverse quark propagator
\begin{eqnarray}
    D^{-1}(p) &=& \begin{pmatrix}
    \slashed{p}-g\phi_0 + \gamma^0\hat{\mu}& ig_\Delta\tau_2\lambda_2\gamma^5\Delta_0\\\
    ig_\Delta\tau_2\lambda_2\gamma^5\Delta_0 & \slashed{p}-g\phi_0-\gamma^0\hat{\mu}
    \end{pmatrix}\;. 
\end{eqnarray}
\end{widetext}
The zeros of the determinant of \(D^{-1}(p)\) give the fermion spectrum. We group the spectrum into the spectrum for the blue quarks
\begin{eqnarray}
\label{un1}
    E_{ub}^{\pm} &=& E\pm\mu_{ub}\;, \\
    E_{db}^{\pm} &=& E\pm\mu_{db}\;, 
\label{un2}
\end{eqnarray}
and the spectrum for the red and green quarks
\begin{eqnarray}
    E_{\Delta^{\pm}}^{\pm} &=& 
    E_{\Delta}^{\pm} \pm \delta\mu
    \;, 
    \label{guppy}    
\end{eqnarray}
where the energies and chemical potentials are defined as
\begin{eqnarray}
E&=&\sqrt{p^2+g^2\phi_0^2}\;,\\
    E^{\pm}_{\Delta} &=& \sqrt{(E{\pm}\bar{\mu})^2+g_\Delta^2\Delta_0^2}\;, 
\\ 
\label{mubar}
\bar{\mu} &=& {1\over2}(\mu_{ur}+\mu_{dg})=
{1\over2}(\mu_{ug}+\mu_{dr})=\mu-{1\over6}\mu_e+{1\over3}\mu_8
\;,
\\
\delta\mu &=& 
 {1\over2}(\mu_{dg}-\mu_{ur})= {1\over2}(\mu_{dr}-\mu_{ug})=
{1\over2}\mu_e
\;,
\end{eqnarray}
The chemical potential $\delta\mu$ is referred to as the mismatch
parameter since it measures the mismatch between the Fermi surfaces
of the $u$ and the $d$ quarks, i.e. the quarks that form pairs.
A large value $\delta\mu$ disfavors Cooper pairing as we shall see below. The ungapped quasiparticles Eqs.~(\ref{un1})--(\ref{un2})
are singlets under the unbroken $SU(2)_c$
symmetry, each with a degeneracy factor one.
The quasiparticles in Eq.~(\ref{guppy})
form two doublets under the unbroken $SU(2)_c$ symmetry, each with 
a degeneracy factor of two. This gives a total of six quark and six
antiquark quasiparticles. They become degenerate when the mismatch
parameter vanishes. The doublet
with energy $E_{\Delta^{+}}^{-}$ has a gap $g_{\Delta}\Delta_0+\delta\mu$, while
the doublet with energy $E_{\Delta^{-}}^{-}$ has a gap $g_\Delta\Delta_0-\delta\mu$ as long as $\delta\mu<g_\Delta\Delta_0$.
When $\delta\mu\geq g_\Delta\Delta_0$, this mode becomes gapless and the
the corresponding phase is referred to as the gapless 2SC phase (g2SC)~\cite{gapless1}.

We now turn to the thermodynamic potential $\Omega$.
We work in the approximation, where bosons are treated at tree level,
whereas we include Gaussian fluctuations of the quarks.
The one-loop contribution to the thermodynamic potential from the quarks is
\begin{widetext}
\begin{eqnarray}
    \Omega_1 &=& -2\int_p\left\{2E+T\log\left[1+e^{-\beta E_{ub}^\pm}\right]
    +T\log\left[1+e^{-\beta E_{db}^\pm}\right]\right\}
    -4\int_p\left\{E_{\Delta}^\pm+T\log\left[1+e^{-\beta E_{\Delta^{\pm}}^\pm}\right]\right\}\;, 
    \label{added}
\end{eqnarray}
which is added to the tree-level term Eq.~(\ref{treepot0}). 
The integrals $\int_p$ are defined in $d=3-2\epsilon$ dimension as
\begin{eqnarray}
\label{dimregdef}
\int_p&=&\left({e^{\gamma_E}\Lambda^2\over4\pi}\right)^{\epsilon}\int{d^dp\over(2\pi)^d}\; ,    
\end{eqnarray}
where $\Lambda$ is the renormalization scale associated with the 
$\overline{\rm MS}$-scheme. The final result for the thermodynamic potential at $T=0$, including quark loops was derived in Ref.~\cite{us} and reads
\begin{eqnarray}
    \Omega_{0+1}&=&
{3\over4}m_{\pi}^2f_{\pi}^2\left\{1-{12m_q^2\over(4\pi)^2f_{\pi}^2}
m_{\pi}^2F^{\prime}(m_{\pi}^2)\right\}{\phi_0^2\over f_{\pi}^2}+{2m_q^4\over(4\pi)^2}\left({9\over2}+\log{\frac{m_q^2}{g_0^2\phi_0^2}} +2\log{\frac{m_q^2}{g_0^2\phi_0^2+g_{\Delta,0}^2\Delta_{0}^2}}\right)\frac{\phi_0^4}{f_\pi^4}\nn\\
&&
\nonumber
-{1\over4}m_{\sigma}^2f_{\pi}^2\left\{1+{12m_q^2\over(4\pi)^2f_{\pi}^2}
\left[\left(1-{4m_q^2\over m_{\sigma}^2}\right)F(m_{\sigma}^2)+{4m_q^2\over m_{\sigma}^2}
-F(m_{\pi}^2)-m_{\pi}^2F^{\prime}(m_{\pi}^2)
\right]
\right\}{\phi_0^2\over f_{\pi}^2}
\\ &&
\nonumber
+{1\over8}m_{\sigma}^2f_{\pi}^2\left\{
1+{12m_q^2\over(4\pi)^2f_{\pi}^2}\left[
\left(1-{4m_q^2\over m_{\sigma}^2}\right)F(m_{\sigma}^2)
-F(m_{\pi}^2)-m_{\pi}^2F^{\prime}(m_{\pi}^2)
\right]
\right\}{\phi_0^4\over f_{\pi}^4}
\\ &&
\nonumber
-{1\over8}m_{\pi}^2f_{\pi}^2\left\{1-{12m_q^2\over(4\pi)^2f_{\pi}^2}
m_{\pi}^2F^{\prime}(m_{\pi}^2)\right\}{\phi_0^4\over f_{\pi}^4}
-m_{\pi}^2f_{\pi}^2\left\{1-{12m_q^2\over(4\pi)^2f_{\pi}^2}m_{\pi}^2F^{\prime}(m_{\pi}^2)\right\}{\phi_0 \over f_{\pi}}
\\ && 
\nonumber
+\left\{m_{\Delta,0}^2-4\bar{\mu}^2\left[1+{4g_{\Delta,0}^2\over(4\pi)^2}\left(\log{m_q^2\over g_0^2\phi_0^2+g_{\Delta,0}^2\Delta_0^2}-F(m_\pi^2)-m_\pi^2F'(m_\pi^2)\right)\right]\right\}\Delta_0^2
\\ &&
\nonumber
+{\lambda_{3,0}\over12}\phi_0^2\Delta_0^2 +{\lambda_{\Delta,0}\over6}\Delta_0^4 + \frac{12g_0^2g_{\Delta,0}^2}{(4\pi)^2}\phi_0^2\Delta_0^2+\frac{6g_{\Delta,0}^4}{(4\pi)^2}\Delta_0^4 
\\ &&
\nonumber
+{4g_{\Delta,0}^4\over (4\pi)^2}\left\{\log{m_q^2\over g_0^2\phi_0^2+g_{\Delta,0}^2\Delta_0^2}-F(m_\pi^2)-m_\pi^2F'(m_\pi^2)\right\}\Delta_0^4
\\
&& 
+{8g_0^2g_{\Delta,0}^2\over (4\pi)^2}\left\{\log{m_q^2\over g_0^2\phi_0^2+g_{\Delta,0}^2\Delta_0^2}-F(m_\pi^2)-m_\pi^2F'(m_\pi^2)\right\}\phi_0^2\Delta_0^2+\Omega^{\rm fin}_1 + \Omega_1^\mu\;, 
\label{eqn:improved}
\end{eqnarray}
where $g_{\Delta,0}^2$ and $\Delta_0^2$ are the running coupling and field
$g_{\Delta,\ms}^2$ and $\Delta_{\ms}^2$ evaluated at a reference scale, but the
product $g_{\Delta,\ms}^2\Delta_{\ms}^2$ is scale invariant. The same remark applies
to $g_0^2$ and $\phi_0^2$.
Moreover, the function $F(p^2)=2-2q\arctan(\mbox{$1\over q$})$    
with $q=\sqrt{{4m_q^2\over p^2}-1}$, and 
\begin{eqnarray}
\Omega_{1}^{\rm fin}&=&-4{\Lambda^{-2\epsilon}}
\int_p\left[
E^{\pm}_{\Delta^{\pm}}-2\sqrt{p^2+g^2\phi_0^2+g_{\Delta}^2\Delta_0^2}
-{\bar{\mu}^2g_{\Delta}^2\Delta_0^2\over(p^2+g^2\phi_0^2+g_{\Delta}^2\Delta^2_0)^{3\over2}}\right]\;,  
\label{numdef1}
\\
    \Omega_1^{\mu} &=& 2{\Lambda^{-2\epsilon}}
    \int_p\left[(E\mp\mu_{ub})\theta(\pm\mu_{ub}-E) + (E\mp\mu_{db})\theta({\pm}\mu_{db}-E)
    +2(E_{\Delta}^{\pm}\mp\delta\mu)\theta({\pm}\delta\mu-E_{\Delta}^{\pm})
\right]
    \;.
\label{eqn:improved2}
\end{eqnarray}
\end{widetext}
Note that the integrals in Eqs.~(\ref{numdef1})--(\ref{eqn:improved2}) are finite in $d=3$. 

\subsection{Nambu-Goldstone bosons in the 2SC phase}
The formation of Cooper pairs and the resulting non-zero gap $\Delta_0$ in the 2SC phase
break the $SU(3)_c$ symmetry down to $SU(2)_c$. In QCD with dynamical
gluons, there is no breaking of the local $SU(3)_c$ symmetry, however,
and the gap $\Delta_0$ is a gauge-variant quantity. Only the magnitude
$|\Delta_0|$ is gauge invariant. Instead of a number of NG modes in the physical spectrum, the Higgs mechanism leads
to five massive gluons, one for each broken generator~\cite{sannino}.
In the QMD model, the $SU(3)_c$ symmetry is global and its breaking to
$SU(2)_c$ should naively lead to five massless excitations, or 
Nambu-Goldstone bosons, corresponding to the five broken generators.~\footnote{The NG bosons have been discussed in the context of the NJL model in Ref.~\cite{GBcounting}. In the NJL model, the NG modes are propagating only after taking quark loops into account.}
Naively, since it is only in Lorentz-invariant theories that Goldstone's
theorem guarantees that the number of massless excitations equals the
number of broken generators. Starting with the paper by Chadha and Nielsen~\cite{holger}, there has been significant progress in the classification of Nambu-Goldstone bosons and their properties; 
see e.g.~\cite{gbcount1,gbcount2,rachel,ata}, and~\cite{brauner} for a comprehensive review.

The complex antitriplet $\Delta_a$ can be parametrized in terms of six real fields.  In order to simplify the discussion and obtain analytical results, we set $\phi_0=0$ in the remainder of this section. There is no loss of generality since the symmetry-breaking pattern in the diquark sector remains the same, $SU(3)_c\rightarrow SU(2)_c$.
Writing $\Delta_3=\Delta_0+\Delta_3=\Delta_0+{1\over\sqrt{2}}(\phi_1+i\phi_2)$
and using a common quark chemical potential $\mu$, the relevant terms up to quadratic order are
\begin{eqnarray}
\nonumber
{\cal L}^{\rm bosons}_{\rm quad}&=&
\nonumber
(\partial_{\mu}+2{i\delta_{\mu0}}\mu)\Delta^{\dagger}_1(\partial_{\mu}-2i\delta_{\mu0}\mu)\Delta_1
\\ &&
\nonumber
{-}(m_{\Delta}^2{+}{\lambda_{\Delta}\over3}\Delta_0^2)\Delta_1^{\dagger}\Delta_1
\\ &&
\nonumber
+(\partial_{\mu}+2{i\delta_{\mu0}}\mu)\Delta^{\dagger}_2(\partial_{\mu}-2{i\delta_{\mu0}}\mu)\Delta_2
\\ &&
\nonumber
-(m_{\Delta}^2{+}{\lambda_{\Delta}\over3}\Delta_0^2)\Delta_2^{\dagger}\Delta_2
\\ &&
\nonumber
+{1\over2}(\partial_{\mu}\phi_1)(\partial^{\mu}\phi_1)
-{1\over2}(m_{\Delta}^2-4{\mu}^2 + \lambda_{\Delta}\Delta_0^2)\phi_1^2
\\ &&
\nonumber
+{1\over2}(\partial_{\mu}\phi_2)(\partial^{\mu}\phi_2)-
{1\over2}(m_{\Delta}^2-4{\mu}^2+{\lambda_{\Delta}\over3}\Delta_0^2)
\phi_2^2
\\ &&
+2{\mu}(\partial_0\phi_1\phi_2-\partial_0\phi_2\phi_1)
\;,
\label{polar}
\end{eqnarray}
The spectrum is found by calculating the zeros of the determinant of the propagator in the usual way. Note that we need to evaluate the
expectation value of the diquark field, $\Delta_0$, 
at the minimum (on shell) of the tree-level thermodynamic potential $\Omega_0$, i.e. for $(m_{\Delta}^2-4{\mu}^2)\Delta_0+{\lambda_{\Delta}\over3}\Delta_0^3=0$. This yields the dispersion relations
\begin{eqnarray}
\label{gab1}
E_{\Delta_1}^{\pm}&=&\sqrt{p^2+4\mu^2}\pm2{\mu}\;,\\
E_{\Delta_2}^{\pm}&=&\sqrt{p^2+4\mu^2}\pm2{\mu}\;,
\label{gab2}
\\
\label{gab3}
E_{\Delta_3}^{\pm}&=&\sqrt{p^2+12{\mu}^2-m_{\Delta}^2\pm\sqrt{16p^2{\mu}^2+(12{\mu}^2-m_{\Delta}^2)^2}}\;.\nonumber \\
&&
\end{eqnarray}
The massless NG modes in Eqs.~(\ref{gab1})--(\ref{gab2}) are 
quadratic for small momenta $p$, while the massless NG mode in 
Eq.~(\ref{gab3}) is linear for small momenta. 
Linear modes are referred to as type A and quadratic modes as type B. Moreover, the partners of the type-B NG modes are massive with mass $4\mu$.

A counting rule for the numbers of type-A and type-B NG modes has been 
derived in Refs.~\cite{gbcount1,gbcount2}. Denoting these numbers by
$n_A$ and $n_B$, respectively, the rule is
\begin{eqnarray}
n_A&=&{\rm dim\,}G-{\rm dim\,}H-{\rm rank\,\varrho}\;,\\
n_B&=&{1\over2}{\rm rank\,}\varrho\;,
\end{eqnarray}
where $G$ is the full symmetry group, $H$ is the unbroken subgroup,
and $\varrho$ is the commutator matrix with matrix elements
\begin{eqnarray}
\label{demat}
\varrho_{ab}&=&\lim_{V\rightarrow\infty}\frac{i}{V}\langle0|[T_a,T_b]|0\rangle\;,
\end{eqnarray}
where $\langle0| A|0\rangle$ is the ground-state expectation value of $A$, and \(V\) is the spatial volume of the system. It is clear that $\varrho_{ab}=0$ whenever the generators $T_a$ and $T_b$ are unbroken. 
In the present case, $G=SU(3)_c$ and $H=SU(2)_c$. The only non-zero matrix elements are $\varrho_{45}=-\varrho_{54}=i\langle[T_4,T_5]\rangle$ and
$\varrho_{67}=-\varrho_{76}=i\langle[T_6,T_7]\rangle$.
This yields ${\rm rank}\,\varrho=4$, and therefore $n_A=1$ and $n_B=2$,
in accordance with our explicit calculations.
The commutators $[T_a,T_b]$ can be written as a linear combination of the symmetry generators $T_3$ and $T_8$. Thus, if color charge neutrality is enforced in the 2SC phase, $Q_8=0$, Eq.~(\ref{demat}) immediately yields $\varrho=0$, and therefore the number of type-A NG bosons, $n_A$, equals the number of broken generators, as first shown in Ref.~\cite{sjefer}. 
In the present case, there are five NG bosons with a linear dispersion relation for small momenta $p$. The massive modes in Eqs.~(\ref{gab1})--(\ref{gab2}) are also special in the sense that their gap is determined by symmetry~\cite{nicolis0}. Their number is given by the number of
pairs of broken generators $T_a$ that do not commute with $T_8$.
There are two such pairs, $T_4\pm iT_5$ and $T_6\pm iT_7$, which leads
to two massive NG bosons. 
Finally, as pointed out in Ref.~\cite{bedaque}, the NG bosons do not Bose condense in the 2SC phase. Condensation corresponds to a change
of the relative flavor orientation of the left and right-handed condensates.
Since these condensates are singlets under $SU(2)_L\times SU(2)_R$, this cannot happen. This is in contrast to NG bosons in the CFL phase in three-flavor QCD~\cite{bedaque2}.


\subsection{Pion condensation at finite isospin $\mu_I$}
In this section, we will discuss charged pion condensation at finite isospin chemical potential and zero baryon chemical potential.
The onset of pion condensation at $T=0$ is a second-order transition
that takes place exactly at the pion mass, $\mu_I=m_{\pi}$.
As mentioned in the Introduction, QCD is amenable to lattice simulations
for $\mu_B$=0, since the sign problem is absent in this case.

Chiral perturbation theory~\cite{chipt,chipt2} provides a systematic low-energy expansion of the chiral Lagrangian
based on the symmetries and degrees of freedom of QCD, and gives model-independent predictions. At finite $\mu_I$, the expansion parameters are
$m_{\pi}^2/(4\pi)^2f_{\pi}^2$ and $\mu_I^2/(4\pi)^2f_{\pi}^2$.
For sufficiently small values of $\mu_I$, we can therefore use 
chiral perturbation theory to describe the BEC phase. The tree-level expressions for the pressure and the energy density 
in the pion-condensed phased are the same for two and three-flavor QCD, and given 
by~\cite{sonstep}
\begin{eqnarray}
p&=&{1\over2}f_{\pi}^2\mu_I^2\left[1-{m_{\pi}^2\over\mu_I^2}\right]^2\;,
\hspace{1cm}m_{\pi}\leq\mu_I\;,
\\
\epsilon(p)&=&-p+2\sqrt{p(p+2f_{\pi}^2m_{\pi}^2)}\;.
\label{eos}    
\end{eqnarray}
In the vacuum phase, which extends from $\mu_I=0$ to $\mu_I=m_{\pi}$,
the pressure and energy density vanish identically, $p=\epsilon=c_s^2=0$.
From the expressions, one can find the speed of sound squared, expressed in terms of the pion mass and the isospin chemical potential,
\begin{eqnarray}
c_s^2&=&{dp\over d\epsilon}=
{\mu_I^4-m_{\pi}^4\over\mu_I^4+3m_{\pi}^4}\;.  
\end{eqnarray}
The speed of sound is zero in the vacuum phase and approaches unity 
in the ultra-relativistic limit $\mu_{I}\rightarrow\infty$.
The large-$\mu_I$ behavior is due to the fact that the pressure is proportional to $\mu_I^2$ and consequently the equation of state becomes $\epsilon=p$ at large $\mu_I$. However, $\chi$PT is not valid for large values of the isospin chemical potential. We can get an idea of the convergence properties of $\chi$PT at finite $\mu_I$ by comparing the predictions for different physical quantities at LO and NLO order.
A conservative estimate suggests that $\chi$PT is reliable for
$\mu_I$ up to 200 MeV~\cite{qing}.
The behavior at large $\mu_I$ is in disagreement with the conformal limit 
of QCD, where $c_s^2={1\over3}$. The reason is that pions are not
the correct degrees of freedom at high (isospin) density. In this region, one should rather think in terms of a Fermi surface that is rendered unstable due to an attractive interaction channel and the formation of loosely bound Cooper pairs of $u$ and $\bar{d}$ quarks. Since Cooper pairs have the same quantum numbers as the pion, one expects a crossover rather than a phase transition. This is called the BEC-BCS crossover.
We will not focus on this crossover, but simply refer to this part of the phase diagram as the BEC/BCS phase.

The quark-meson model has also been used as a low-energy model to study
QCD at finite isospin~\cite{hepion,smek,kneschke,ayala,kojo,farias,scoc,brandtqm}. Denoting the pion condensate by $\rho_0$, the tree-level thermodynamic potential is
\begin{eqnarray}
\nonumber
\Omega_0&=&{1\over2}m^2\phi_0^2+{1\over2}(m^2-\mu_I^2)\rho_0^2
+{\lambda\over24}(\phi_0^2+\rho_0^2)^2-h\phi_0\;.
\\ &&
\label{treepot}
\end{eqnarray}
The one-loop thermodynamic potential with on-shell renormalization was derived in Ref.~\cite{kneschke} for $N_c$ colors, and we give it here for completeness,
\begin{widetext}
\begin{eqnarray}
\nonumber
\Omega_{0+1}&=&
\dfrac{3}{4}m_\pi^2 f_\pi^2
\left\{1-\dfrac{4 m_q^2N_c}{(4\pi)^2f_\pi^2}m_\pi^2F^{\prime}(m_\pi^2)
\right\}\dfrac{\phi_0^2+\rho_0^2}{f_{\pi}^2}
\\ \nonumber &&
 -\dfrac{1}{4}m_\sigma^2 f_\pi^2
\left\{
1 +\dfrac{4 m_q^2N_c}{(4\pi)^2f_\pi^2}
\left[ \left(1-\mbox{$4m_q^2\over m_\sigma^2$}
\right)F(m_\sigma^2)
 +\dfrac{4m_q^2}{m_\sigma^2}
-F(m_\pi^2)-m_\pi^2F^{\prime}(m_\pi^2)
\right]\right\}\dfrac{\phi_0^2+\rho_0^2}{f_{\pi}^2}
\\ \nonumber &&
-\frac{1}{2}\mu_I^2f_\pi^2
\left\{1-\dfrac{4 m_q^2N_c}{(4\pi)^2f_\pi^2}
\left[\log\mbox{$\phi_0^2+\rho_0^2\over f_{\pi}^2$}
+F(m_\pi^2)+m_\pi^2F^{\prime}(m_\pi^2)\right]
\right\}{\rho_0^2\over f_{\pi}^2}
\\ \nonumber
 & & + \dfrac{1}{8}m_\sigma^2 f_\pi^2
\left\{ 1 -\dfrac{4 m_q^2  N_c}{(4\pi)^2f_\pi^2}\left[
\dfrac{4m_q^2}{m_\sigma^2}
\left( 
\log\mbox{$\phi_0^2+\rho_0^2\over f_{\pi}^2$}
-\mbox{$3\over2$}
\right) -\left( 1 -\mbox{$4m_q^2\over m_\sigma^2$}\right)F(m_\sigma^2)
+F(m_\pi^2)+m_\pi^2F^{\prime}(m_\pi^2)\right]
 \right\}\dfrac{(\phi_0^2+\rho_0^2)^2}{f_{\pi}^4}
\\ &&
- \dfrac{1}{8}m_\pi^2 f_\pi^2
\left[1-\dfrac{4 m_q^2N_c}{(4\pi)^2f_\pi^2}m_\pi^2F^{\prime}(m_\pi^2)\right]
\dfrac{(\phi_0^2+\rho_0^2)^2}{f_{\pi}^4}
-m_\pi^2f_\pi^2\left[
1-\dfrac{4 m_q^2  N_c}{(4\pi)^2f_\pi^2}m_\pi^2F^{\prime}(m_\pi^2)
\right]\dfrac{\phi_0}{f_{\pi}}
+\Omega_{1}^{\rm fin}
+\Omega_1^{\mu}
\;,
\label{fullb}
\end{eqnarray}
where $\rho_0^2=\rho^2_{0,\ms}$ at a reference scale, with $g_0^2\rho_0^2=g_{0,\ms}^2\rho_{0,\ms}^2$ being scale invariant, and 
\begin{eqnarray}
\label{fd0}
\Omega_{1}^{\rm fin}&=&-2N_c\Lambda^{-2\epsilon}\int_p\left[E_{\rho_0}^{\pm}-2\sqrt{p^2+g_0^2(\phi_0^2+\rho_0^2)}
-{\mu_I^2\rho_0^2\over4[p^2+g_0^2(\phi_0^2+\rho_0^2)]^{3\over2}}
\right]\;,\\
\Omega_1^{\mu}&=&
2N_c\Lambda^{-2\epsilon}\int_p\left[(E_{\rho_0}^{\pm} - \mu)\theta(\mu-E_{\rho_0}^{\pm}) +(E_{\rho_0}^{\pm}+\mu)\theta(-\mu-E_{\rho_0}^{\pm})\right]\;.
\label{fd}
\end{eqnarray}
\end{widetext}
Here the fermionic quasiparticle spectrum is
\begin{eqnarray}
E_{\rho_0}^{\pm}&=&\sqrt{(E\pm\mbox{$1\over2$}\mu_I)^2
+g_0^2\rho_0^2}\;,    
\label{piongap}
\end{eqnarray}
with $E=\sqrt{p^2+g_0^2\phi_0^2}$.
Note the similarity between the gapped spectra Eq.~(\ref{guppy})
and Eq.~(\ref{piongap}) for $\delta\mu=0$. 
The integral Eqs.~(\ref{fd0})
is evaluated numerically directly in $d=3$ dimensions.
Eqs.~(\ref{eqn:improved}) and~(\ref{fullb}) is the starting point for mapping out the phase diagram.

\subsection{Behavior at large chemical potentials}
\label{largebe}
For large chemical potentials $\mu$ or $\mu_I$, we can obtain simple expressions for thermodynamic quantities such as pressure and energy density. We first consider large $\mu$.

\begin{widetext}
If we are deep in the 2SC phase, we have $\phi_0\ll\Delta_0$.
The thermodynamical potential in Eq.~(\ref{eqn:improved}) reduces to
\begin{eqnarray}
\nonumber
\Omega_{0+1}&=&
(m_{\Delta,0}^2-4\bar{\mu}^2)\Delta_0^2+{\lambda_{\Delta,0}\over6}\Delta_0^4
-{4(4\bar{\mu}^4+{\mu}_{ub}^4+{\mu}_{db}^4)\over3(4\pi)^2}   
-{16g_{\Delta,0}^2\over(4\pi)^2}\left[\log{m_q^2\over g_{\Delta,0}^2\Delta_0^2}
-F(m_{\pi}^2)-m_{\pi}^2F^{\prime}(m_{\pi}^2)
\right]\bar{\mu}^2\Delta_0^2
\nonumber
\\ &&
+{4g_{\Delta,0}^4\over (4\pi)^2}\left[\log{m_q^2\over g_{\Delta,0}^2\Delta_0^2}+{3\over2}-F(m_{\pi}^2)-m_{\pi}^2F^{\prime}(m_{\pi}^2)\right]\Delta_0^4\; .
\end{eqnarray}
In the same approximation, the gap equation reads
\begin{eqnarray}
\nonumber
0&=&m_{\Delta,0}^2-4\bar{\mu}^2
+{\lambda_{\Delta,0}\over3}\Delta_0^2
-{16g_{\Delta,0}^2\over(4\pi)^2}\left[\log{m_q^2\over g_{\Delta,0}^2\Delta_0^2}-1
-F(m_{\pi}^2)-m_{\pi}^2F^{\prime}(m_{\pi}^2)
\right]\bar{\mu}^2
\\ &&
+{8g_{\Delta,0}^4\over (4\pi)^2}\left[\log{m_q^2\over g_{\Delta,0}^2\Delta_0^2}+1
-F(m_{\pi}^2)-m_{\pi}^2F^{\prime}(m_{\pi}^2)
\right]\Delta_0^2\;. 
\label{eqn_gap_delta}
\end{eqnarray}
\end{widetext}
Assuming $\bar{\mu}\gg\Delta_0$, we can solve the gap equation explicitly.
Denoting $\Delta_0$ by $\bar{\Delta}_0$ in this regime, we find
\begin{eqnarray}
\bar{\Delta}_0^2&=&{m_q^2\over g_{\Delta,0}^2}\exp\left[{{(4\pi)^2\over4g_{\Delta,0}^2}-1-F(m_{\pi}^2)-m_{\pi}^2F^{\prime}(m_{\pi}^2)}\right]\;. 
\label{gapleik}
\end{eqnarray}
In other words, the gap $g_{\Delta,0}\Delta_0$ approaches a constant as $\bar{\mu}\rightarrow\infty$. This is in contrast to the behavior of the gap at tree-level. In this case, the gap increases linearly with $\bar{\mu}$. The difference is caused by the logarithms introduced by adding quark loops.
Note that the gap $g_{\Delta,0}\bar{\Delta}_0$ depends only
on the diquark coupling $g_{\Delta,0}$ and the zero-temperature 
quark mass $m_q$. In principle, we can use the result Eq.~(\ref{gapleik}) to connect to perturbative QCD. We simply tune the parameter so that it
agrees with the gap calculated in pQCD for a given large value of 
$\mu$.~\footnote{We have to multiply by \(2^{-{1\over3}}\)
since the CFL gap is related to the 2SC gap by this factor~\cite{alford}.}

The pressure $p$ is given by minus the thermodynamic potential $\Omega_{0+1}$, evaluated at the solution of the gap equation.
At high densities, the pressure reduces to
\begin{eqnarray}
\nonumber
p&=&
{4(4\bar{\mu}^4+{\mu}_{ub}^4+{\mu}_{db}^4)\over3(4\pi)^2}   
+{16\bar{\mu}^2g_{\Delta,0}^2\bar{\Delta}_0^2 \over(4\pi)^2}
+{4\mu_e^4\over3(4\pi)^2}
\;.
\\
\label{lp}
\end{eqnarray}
Except for the term ${4(\mu_{sr}^4+\mu^4_{sg}+\mu_{sb}^4)\over3(4\pi)^2}$
coming from unpaired $s$ quarks, the result for the pressure Eq.~(\ref{lp}) is the same as obtained for the 2SC phase in
the three-flavor case based on an analytic approximation of the NJL model
~\cite{steinerneutral}.~\footnote{The chemical potentials $\mu_{sr}$, $\mu_{sg}$, and $\mu_{sb}$ are expressed in terms of $\mu$, $\mu_e$, $\mu_8$ as well as the mass $m_s$ of the $s$ quark.}
The corrections to the pressure due to the color superconductivity are
parametrically suppressed by $\bar{\Delta}_0^2/\bar{\mu}^2$, which can be understood as follows. Whereas Cooper pairing affects only states close to the Fermi surface, the Pauli pressure receives contributions from the entire Fermi sphere. The energy density $\epsilon$ and trace anomaly
$\langle T_{\,\,\alpha}^{\alpha}\rangle$ are given by
\begin{eqnarray}
\label{le}
\epsilon&=&
{4(4\bar{\mu}^4+{\mu}_{ub}^4+{\mu}_{db}^4)\over(4\pi)^2} 
+{16\bar{\mu}^2g_{\Delta,0}^2\bar{\Delta}_0^2\over(4\pi)^2}
+{4\mu_e^4\over(4\pi)^2}
\;,\\
\langle T_{\,\,\alpha}^{\alpha}\rangle&=&-{32\bar{\mu}^2g_{\Delta,0}^2\bar{\Delta}_0^2\over(4\pi)^2}\;.
\end{eqnarray}
By imposing neutrality constraints, we can also obtain simple expressions for 
\(\mu_e\) and \(\mu_8\) for very large values of $\mu$.
Expanding the thermodynamic potential around this asymptotic value
and solving for the neutrality constraints Eq.~\eqref{neucond}, we arrive at the following expressions through order ${1\over\mu}$
\begin{eqnarray}
  \nonumber
    \mu_e &=& \frac{3}{11}\left[1+\left({9\over2}\right)^{2\over3}-\left({9\over2}\right)^{1\over3}\right]\mu
    \\ &&
    \nonumber
\times\left\{1 + {2\over 9}
\left[1+\left({2\over9}\right)^{1\over3}\right]
\frac{g_{\Delta,0}^2\bar{\Delta}_0^2}{\mu^2}\right\}
    \\ &\approx&
        \label{eqn:mueasymptotic}
        0.57\mu\left[1+0.36
\frac{g_{\Delta,0}^2\bar{\Delta}_0^2}{\mu}\right]
\;,\\
    \nonumber
    \mu_8 &=& \frac{9}{11}\left[1-\left({375\over32}\right)^{1\over3}+\left({125\over48}\right)^{1\over3}
    \right]\mu
    \\ &&
\times\left\{1
-{50\over27}\left[1+\left({162\over125}\right)^{1\over3}+\left({4\over3}\right)^{1\over3}\right]
    \frac{g_{\Delta,0}^2\bar{\Delta}_0^2}{\mu^2} \right\}
    \nonumber
    \\
    &\approx&
            0.09\mu\left[1-5.91
\frac{g_{\Delta,0}^2\bar{\Delta}_0^2}{\mu}\right]
\;.
    \label{eqn:mu8asymptotic}
\end{eqnarray}
It is interesting to note that both chemical potentials $\mu_e$ and $\mu_8$
are independent of the gap at asymptotic values of the quark chemical potential and therefore grow linearly with \(\mu\). This is in contrast to the 2SC phase with three flavors~\cite{steinerneutral}. The pressure $p$ receives the extra contribution from the unpaired $s$ quarks as explained below Eq.~(\ref{lp}). 
The extra contributions to the neutrality equations from this term
cancel the leading-order contributions in 
Eqs.~(\ref{eqn:mueasymptotic})--(\ref{eqn:mu8asymptotic})
(for $m_s=0$) so that $\mu_e$ and $\mu_8$ are both of order ${\bar{\Delta}_0^2\over\mu}$.
Inserting the expansions Eqs.~(\ref{eqn:mueasymptotic}) and~(\ref{eqn:mu8asymptotic})
into the expressions for the pressure Eq.~(\ref{lp}) and energy density Eq.~(\ref{le}),
we obtain the speed of sound for neutral matter at large $\mu$
\begin{eqnarray}
\nonumber
c_s^2 &=& 
{1\over3}\left\{1+{4\over9}\left[1+\left({2\over9}\right)^{1\over3}\right]
\frac{g_{\Delta,0}^2\bar{\Delta}^2}{\mu^2}\right\}
\\
    \label{speed2sc}
    &\approx& \frac{1}{3}\left[1+0.71\frac{g_{\Delta,0}^2\bar{\Delta}_0^2}{\mu^2}\right]\;.
    \end{eqnarray}
In this form, it is easy to see that the speed of sound approaches the conformal limit from above. In the case $\mu_e=\mu_8=0$, the coefficient of the correction
term is ${2\over3}$~\cite{us}. Thus, the speed of sound relaxes faster to ${c_s}={1\over\sqrt{3}}$ if charge neutrality is not imposed.

Comparing Eqs.~(\ref{eqn:improved}) and~(\ref{fullb}), we note that the expressions for the thermodynamic potential are essentially the same, the one-loop terms differ by numerical factors and the relevant couplings. This implies that the gap, pressure, energy density, trace anomaly, and speed of sound have the same behavior at large $\mu_I$. As in the 2SC phase, we calculate the thermodynamic potential deep into the BCS
phase using $\phi_0\approx0$. This yields
\begin{eqnarray}
\bar{\rho}_0^2&=&
{m_q^2\over g_{0}^2}\exp\left[{{(4\pi)^2\over4{N_c}g_{0}^2}-1-F(m_{\pi}^2)-m_{\pi}^2F^{\prime}(m_{\pi}^2)}\right]
\;,\\
p&=&
{N_c{\mu}_I^4\over{6}(4\pi)^2}+{{2N_c}{\mu}_I^2g_{0}^2\bar{\rho}_0^2\over(4\pi)^2}\;,\\ 
\epsilon&=&{{N_c}{\mu}_I^4\over2(4\pi)^2}+{{2N_c}{\mu}_I^2g_{0}^2\bar{\rho}_0^2\over(4\pi)^2}\;,\\
\langle T_{\,\,\alpha}^{\alpha}\rangle&=&-
{{4N_c}{\mu}_I^2g_{0}^2\bar{\rho}_0^2\over(4\pi)^2}\;,\\
c_s^2 &=& 
    {1\over3}\left[1+{4g_{0}^2\bar{\rho}_0^2\over\mu_I^2}\right]\;.
\label{speed}
\end{eqnarray}
In contrast to the superconducting gap, the asymptotic value of the pion condensate $\bar{\rho}_0$ is completely determined by vacuum physics.

\section{Results and discussion}
In this section, we discuss our numerical results. 
The values of the meson masses and the pion decay constant are
\begin{eqnarray}
m_{\pi}&=&140\,\text{MeV}\;, \\
m_{\sigma}&=&600\,\text{MeV}\;, \\
f_{\pi}&=&93\,\text{MeV}\;. 
\end{eqnarray}
In addition, the quark mass is
\begin{eqnarray}
m_q&=&300\,\text{MeV}\;. 
\end{eqnarray}
We also set $N_c=3$. 
As we have explained, the remaining parameters are treated as free.
A natural guess for the parameters would be $g_{\Delta}\sim g$, 
$\lambda\sim\lambda_3\sim\lambda_{\Delta}$, which in~\cite{us}
was shown to yield reasonable results. It turns out that the results are sensitive mainly to the diquark mass parameter $m_{\Delta}$ and the diquark coupling $g_{\Delta}$. The position of the transition to the 2SC phase depends on both parameters, while the gap is largely determined by
the latter. For asymptotically large values of $\mu$, the gap depends only
on $\bar{\Delta}_0$ (in addition to the quark and pion mass).

To illustrate the possible phase diagrams, we have chosen two sets of parameters,
\begin{eqnarray}
{\rm set\,1:}&&
\hspace{0.7cm}
m_{\Delta}={500\;\rm MeV}\;,
\hspace{0.7cm}
g_{\Delta}= 2g\;,\\
&&\hspace{0.7cm} \lambda_3=\lambda_0\;,
\hspace{0.7cm} \lambda_\Delta=\lambda_0/4\;,\\
{\rm set\,2:}&&
\hspace{0.7cm}
m_{\Delta}={900\;\rm MeV}\;,
\hspace{0.7cm}
g_{\Delta}=1.5g\;,\\
&&\hspace{0.7cm} \lambda_3=0\;, \hspace{0.7cm} \lambda_\Delta=\lambda_0/5\;.
\end{eqnarray}
The first set is one of the four parameter sets used in our previous analysis in Ref.~\cite{us}. The second set was chosen to show that NQM is possible when neutrality has been imposed (see Fig.~\ref{condi2} below).
We mentioned in Sec.~\ref{largebe} that we can determine $g_{\Delta}$ by equating our asymptotic value for the gap to the gap at large densities obtained from pQCD. We can also determine it from other constraints on the gap at large densities.
Recently,~\cite{rajakurk} used astrophysical observations to obtain an upper bound on the CFL gap of 216 MeV at baryon chemical potential $\mu_B=2.6$ GeV.
We therefore ensure that our gap for the first set does not exceed this
limit \(g_\Delta\Delta_{0}\leq 268=2^{-{1\over3}}216 \) MeV for large \(\mu\) ,while the gap for set two we allow to go beyond. More generally, the parameter with the largest effect is
\(g_\Delta\). Increasing \(g_\Delta\) leads to an earlier transition from vacuum or NQM to the 2SC phase. Increasing \(g_\Delta\) also leads to lower gap values. Constraining this parameter is therefore most important. 
Lastly, increasing (decreasing) \(m_\Delta\) leads to a later (earlier) transition to the 2SC phase, either from the vacuum or NQM.

To find the minimum of the thermodynamic potential $\Omega_{0+1}$, we must simultaneously solve the two gap equations
\begin{eqnarray}
    \frac{\partial \Omega_{0+1}}{\partial\phi_0} &=&  \frac{\partial \Omega_{0+1}}{\partial\Delta_0} =0\; ,
\hspace{2mm}({\rm 2SC})
\label{mini}
    \\
    \frac{\partial \Omega_{0+1}}{\partial\phi_0} &=&  \frac{\partial \Omega_{0+1}}{\partial\rho_0} =0\;,  
 \hspace{2mm}   
    ({\rm BEC/BCS})
\end{eqnarray}
using Eq.~(\ref{eqn:improved}) and Eq.~(\ref{fullb}), respectively, 
and then choose the global minimum. Imposing electric charge and color charge neutrality, the Eqs.~(\ref{neucond}) and~(\ref{mini}) are solved simultaneously.

\begin{figure}[htb!]
    \centering
    \includegraphics[width=\linewidth]{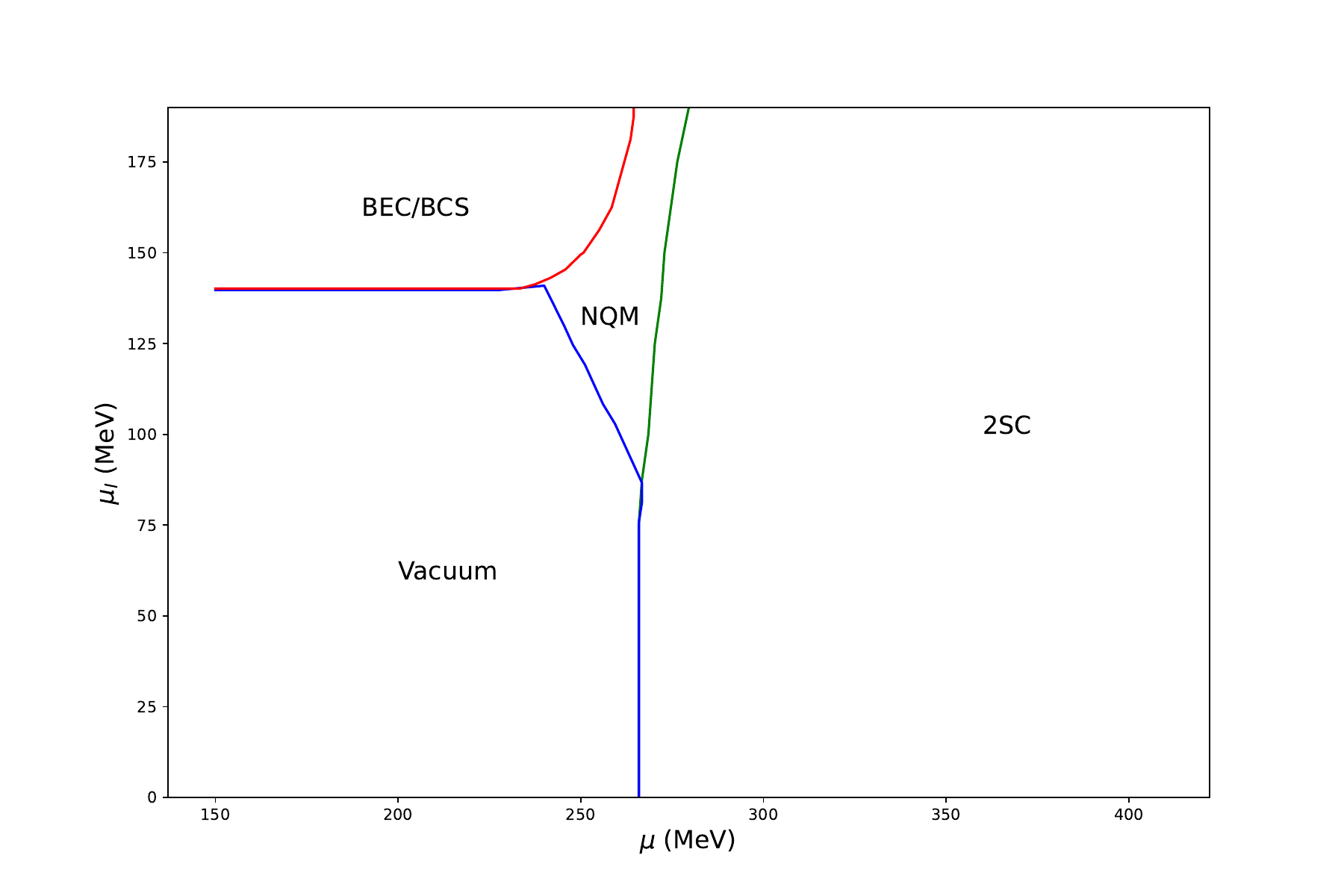}
    \caption{Phase diagram in the $\mu$--$\mu_I$ plane at $T=0$ for parameters set 1. See main text for details.}
    \label{diagram}
\end{figure}

In Fig.~\ref{diagram}, we show the zero-temperature phase diagram
in the $\mu$--$\mu_I$ plane using parameter set 1. The red line is the phase boundary between the pion-condensed phase and the vacuum phase or normal quark matter. The blue line is the phase boundary between the vacuum phase and either normal quark matter or the 2SC phase, whereas the green line is the phase boundary between NQM  and the color superconducting phase.

\begin{figure}[htb!]
    \centering
    \includegraphics[width=\linewidth]{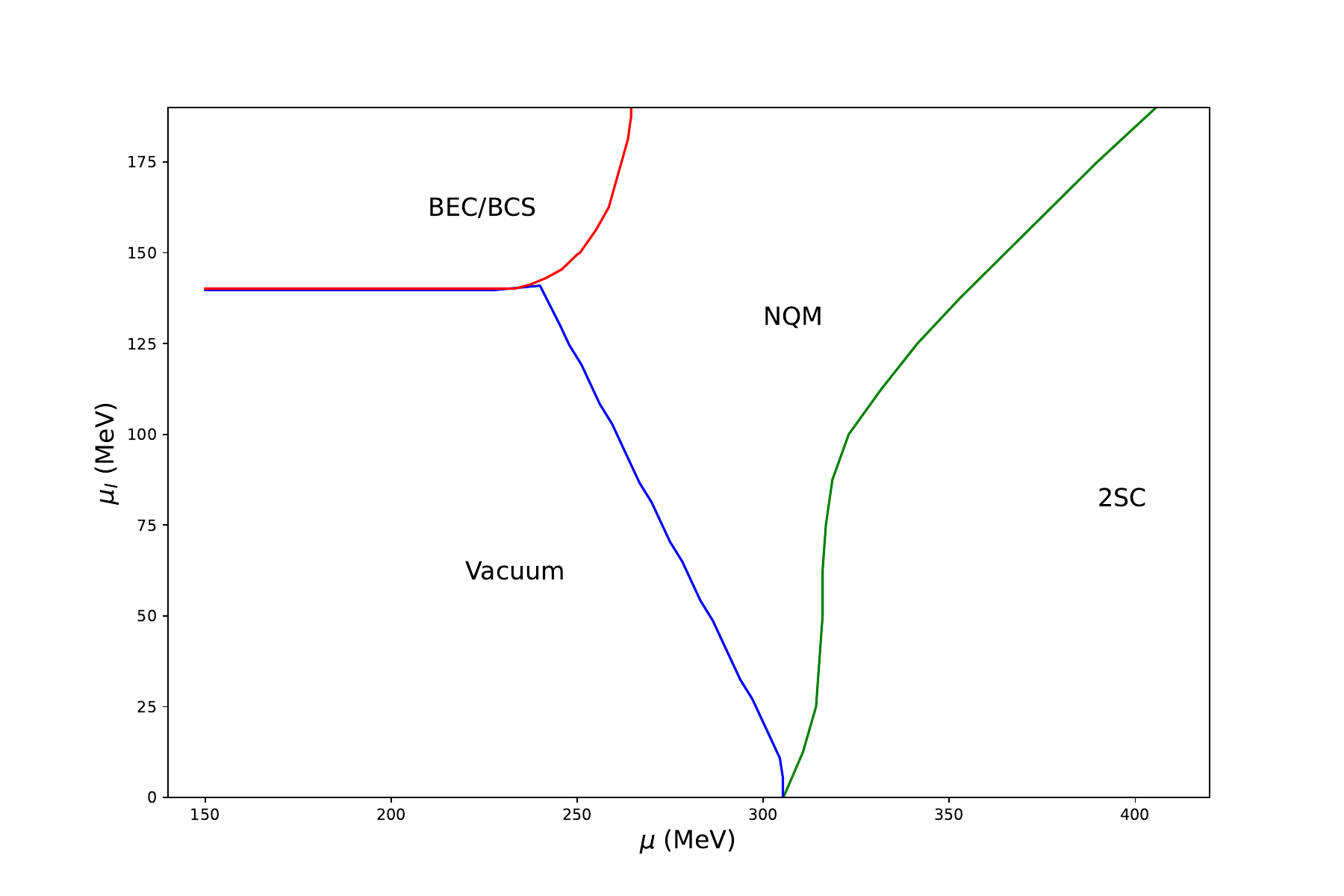}
    \caption{Phase diagram in the $\mu$--$\mu_I$ plane at $T=0$ for parameters set 2.
    See main text for details.}
    \label{diagram2}
\end{figure}

In Fig.~\ref{diagram2}, we show the zero-temperature phase diagram
in the $\mu$--$\mu_I$ plane using parameter set 2. The color coding
is the same as in Fig.~\ref{diagram}, but now normal quark matter exists
in a much larger region in the plane. Below we shall see that this gives rise to electrically neutral normal quark matter.

\begin{figure}[htb!]
    \centering
    \includegraphics[width=\linewidth]{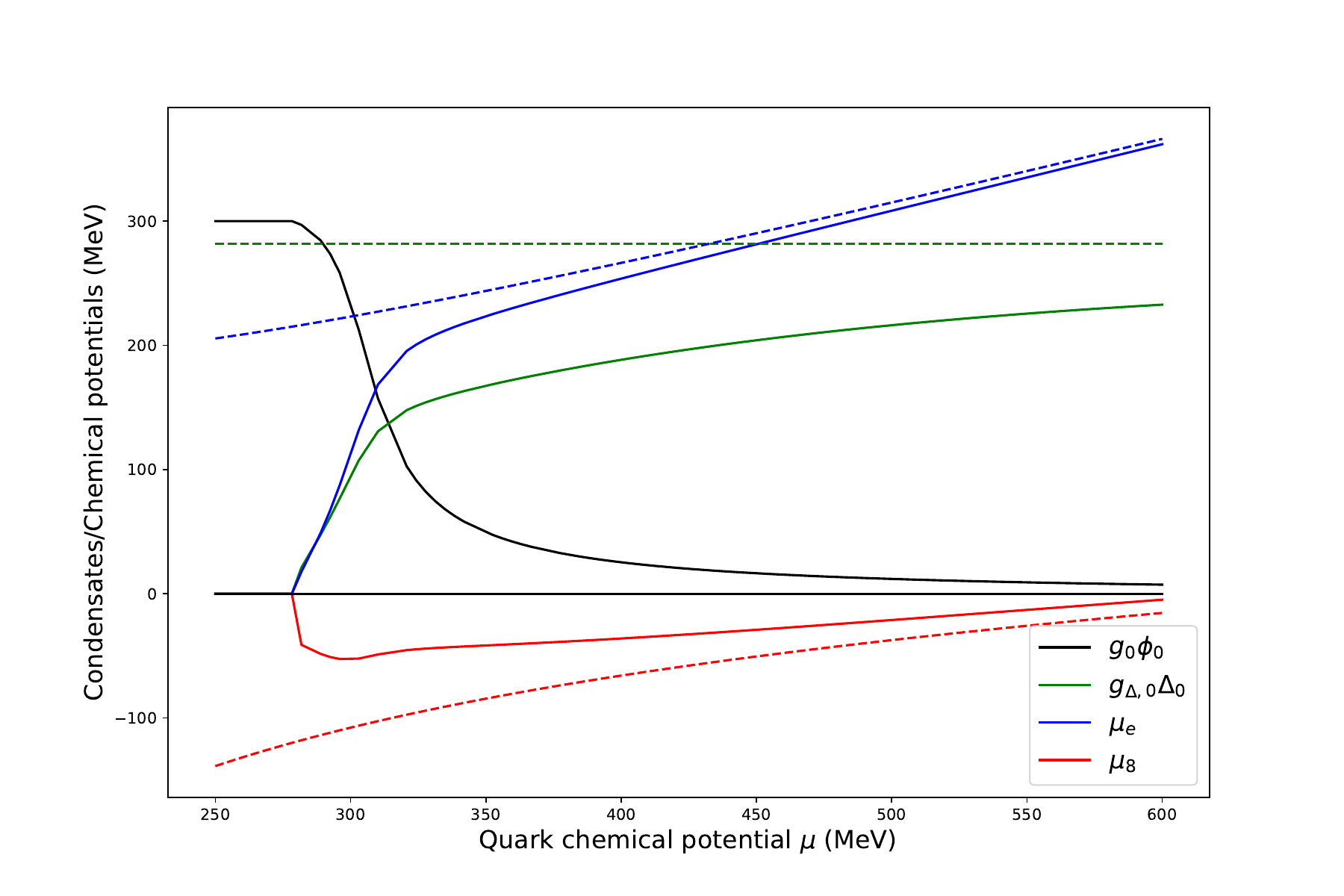}
\caption{Quark condensate (black line), superconducting gap (green line),
electron chemical potential (blue line), and color chemical potential (red line) as functions of ${\mu}$ for parameter set 1. See main text for details.}
    \label{condi}
\end{figure}

In Fig.~\ref{condi}, we show the quark condensate (black line), 
the superconducting gap (green line), the electron chemical potential $\mu_e$ (blue line) and the color chemical potential $\mu_8$ (red line) as functions of the quark chemical potential $\mu$ for parameter set 1. For these parameters, the transition is directly from the vacuum to the 2SC phase. 
The chemical potentials vanish in the entire region up to the transition to the 2SC phase, reflecting that the vacuum state is neutral.
Since the 2SC ground state carries both color and electric charge, both chemical potentials 
$\mu_e$ and $\mu_8$ are non-zero in order to satisfy the charge neutrality conditions. Note that 
$\mu_e>0$ such that the electron density is positive, which is needed to neutralize the positively charged 2SC state. The superconducting gap approaches a constant
for large values of the chemical potential, cf Eq.~(\ref{gapleik}) and the green dashed line in the Fig.
This is in contrast to the NJL model
regulated by a conventional three-dimensional cutoff $\Lambda$, where the gap has a maximum at $\mu\approx500$ MeV, after which it drops rather quickly to zero.
As pointed out in Ref.~\cite{huangneutral}, this unphysical behavior
is a regularization artifact. However, in recent work on the NJL model using
renormalization group ideas~\cite{gholami}, this problem is absent.

In Fig.~\ref{condi2}, we show the same quantities as in Fig.~\ref{condi},
but now for parameter set 2. We also show the asymptotic value for the gap (green dashed line), the asymptotic behavior of $\mu_e$ (dashed blue line) and 
of $\mu_8$ (dashed red line), 
cf. Eqs.~(\ref{eqn:mueasymptotic})--(\ref{eqn:mu8asymptotic}).
The difference is that there is a phase of normal quark matter between the vacuum phase and the 2SC phase.
Two-flavor normal quark matter is not electrically neutral without an
electron background, which is why $\delta\mu={1\over2}\mu_e$ (blue line) becomes 
nonzero before the superconducting gap (green line) and $\mu_8$ (red line). 
Note that the red lines cross the $y$-axis, so $\mu_8$ becomes positive for sufficiently large values of $\mu$, which is in agreement with large-$\mu$ behavior 
given by (\ref{eqn:mu8asymptotic}).

\begin{figure}[htb!]
    \centering
    \includegraphics[width=\linewidth]{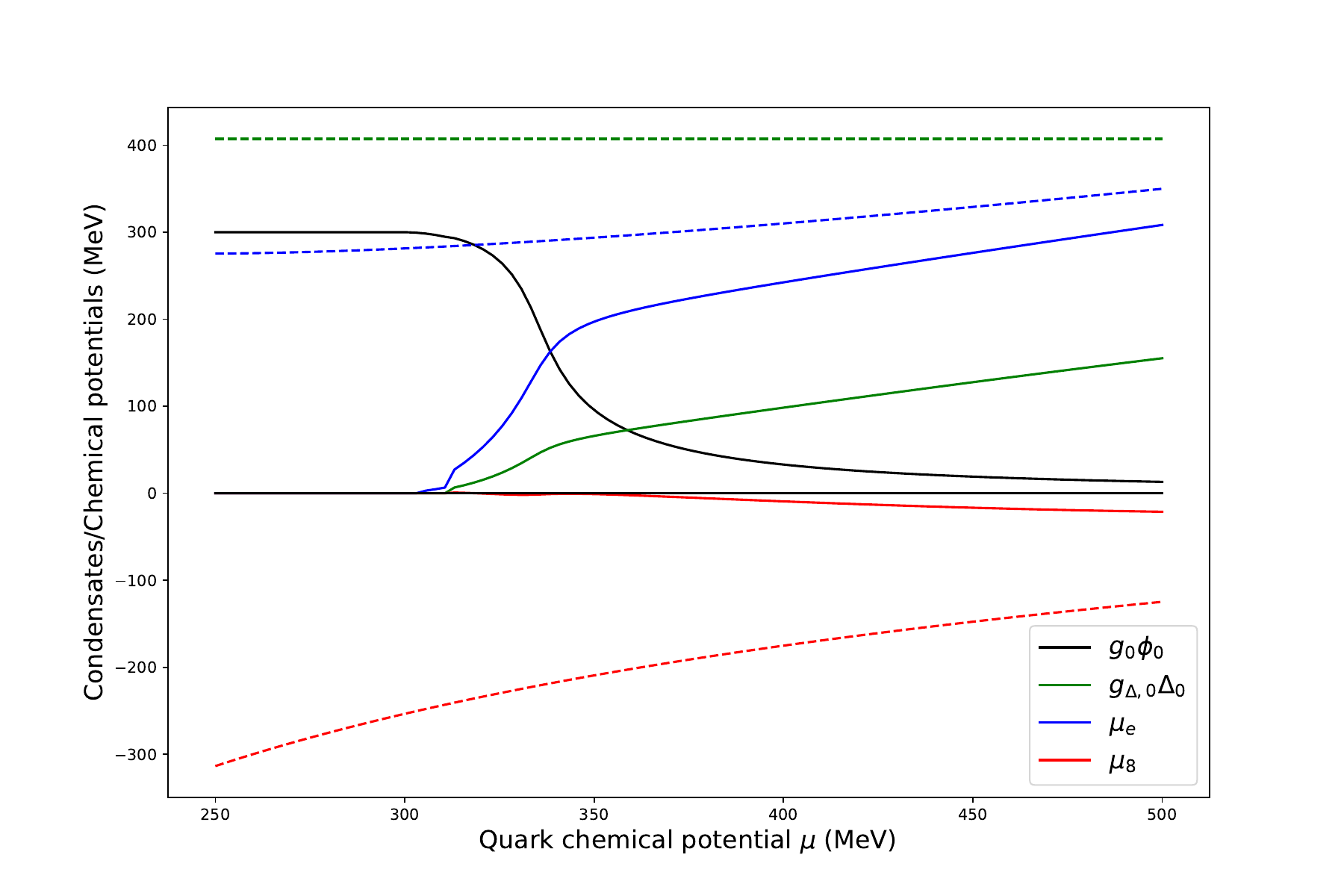}
\caption{Quark condensate (black line), superconducting gap (green line),
electron chemical potential (blue line), and color chemical potential (red line) as functions of ${\mu}$ for parameter set 2. See main text for details.}
    \label{condi2}
\end{figure}

\begin{figure}[htb!]
    \centering
    \includegraphics[width=\linewidth]{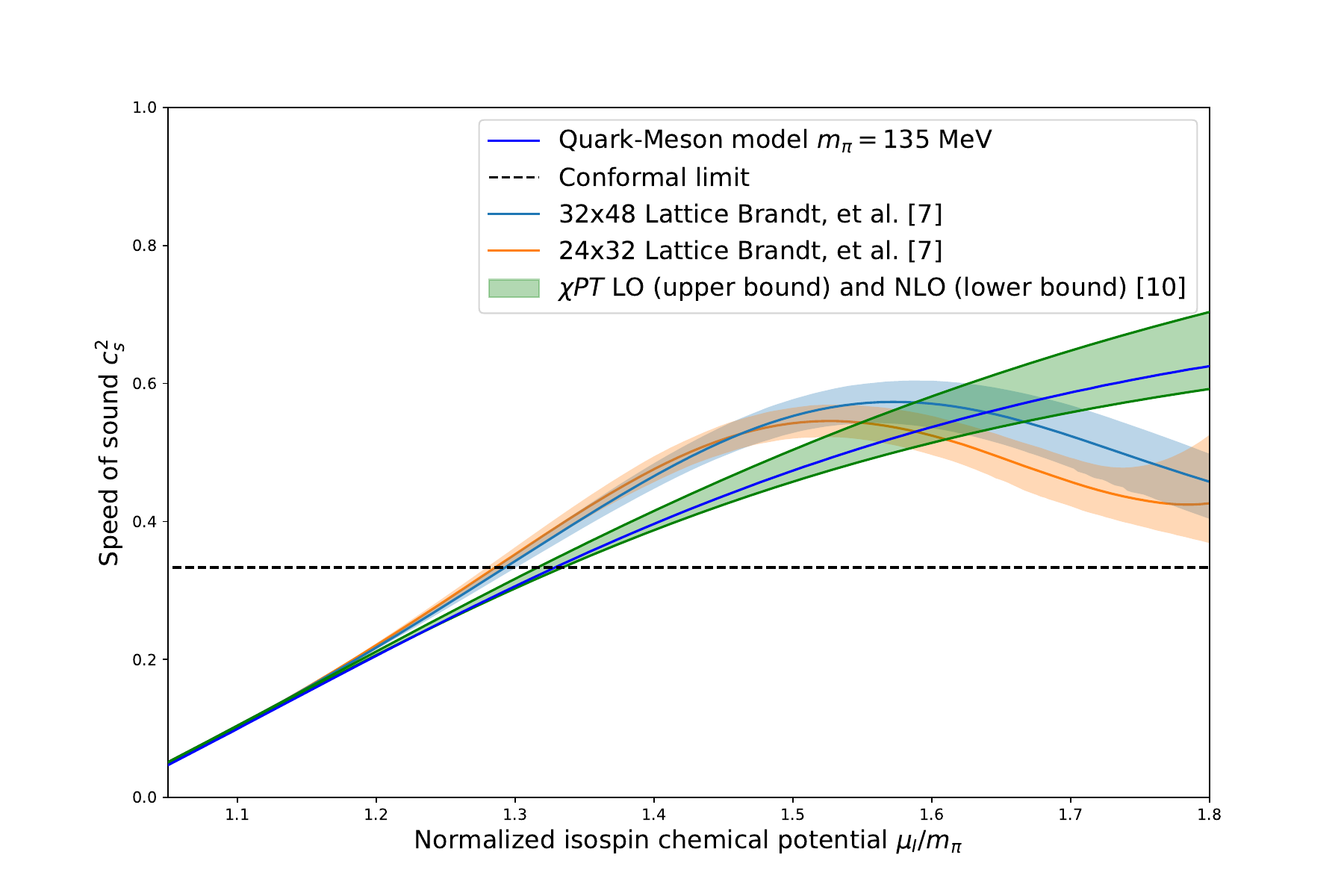}
    \caption{Speed of sound squared in the pion-condensed phase versus $\mu_I/m_{\pi}$. See main text for details.}
   \label{speedoiso}
\end{figure}

\begin{figure}[htb!]
    \centering
    \includegraphics[width=\linewidth]{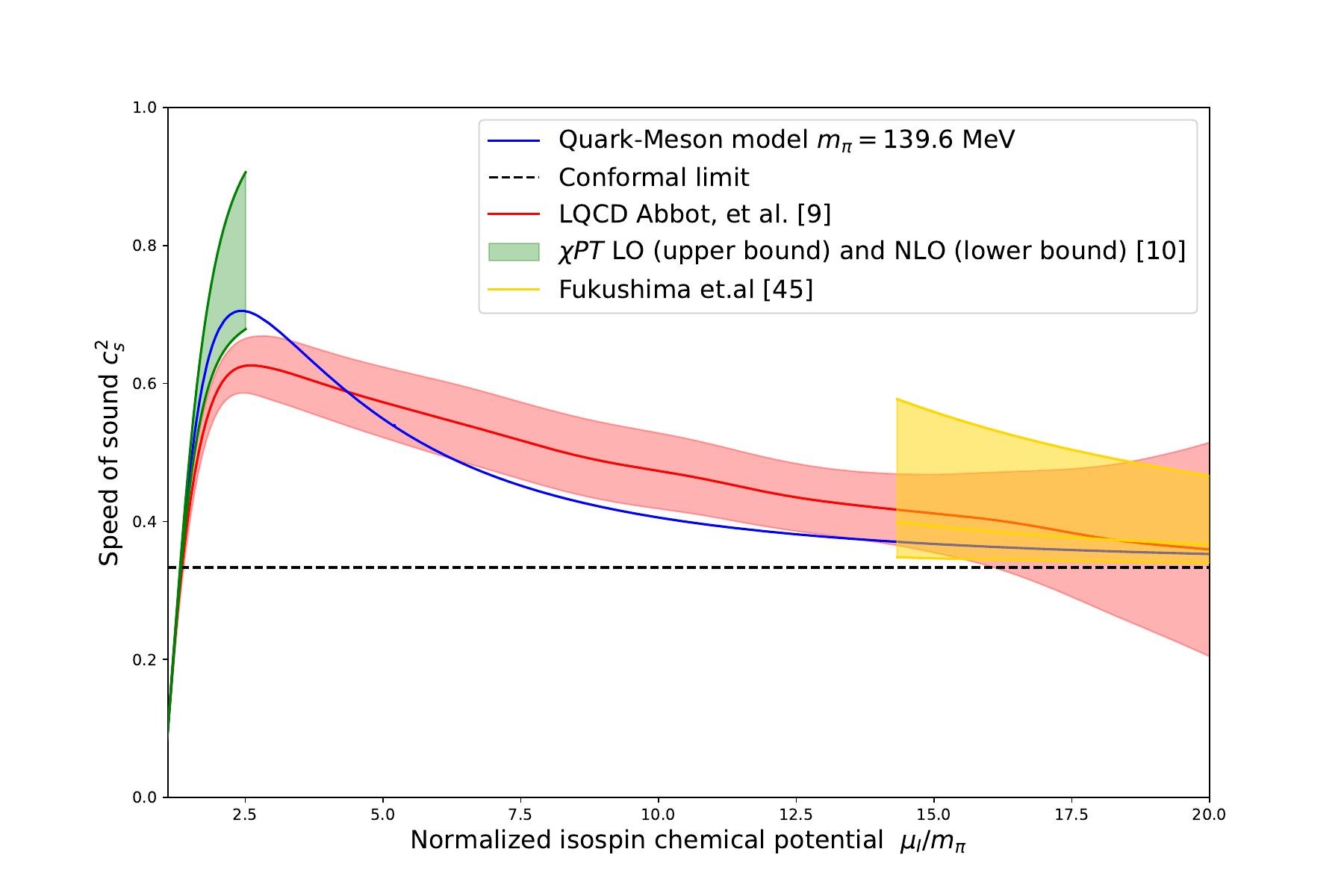}
    \caption{Speed of sound squared in the pion-condensed phase versus $\mu_I/m_{\pi}$. The pion mass is \(m_\pi=139.6\) MeV from \cite{abb24}. See main text for details.}
    \label{speedoiso2}
\end{figure}

Next, we consider the speed of sound, which is of interest because it gives information about the stiffness of the equation of state. The EoS is crucial in understanding
the structure of neutron stars. Before we discuss the case of finite quark chemical potential, we consider finite isospin for which lattice data exist. For similar results and analyses, see Ref.~\cite{kojo} and very recently Ref.~\cite{brandtqm}. 
In Fig.~\ref{speedoiso}, we show the result for the speed of sound squared $c_s^2$ as a function of $\mu_I/m_{\pi}$ using the quark-meson model (blue line).
We also show the results for two-flavor chiral perturbation theory~\cite{qing}
to leading-order (LO) and next-to-leading-order (NLO)  The green band is simply the region between the LO (upper line) and the NLO (lower line) results. Note that the blue curve lies entirely within this band. We also show the results of lattice simulations using two different lattices~\cite{lattice}, given by the light blue and yellow bands.
In the lattice simulations, $m_{\pi}=135$ MeV and $f_{\pi}=90$ MeV were used.  
In this range of isospin chemical potentials, $\chi$PT and the
quark-meson model are in good qualitative agreement with the lattice data. although the peak in $c_s^2$
occurs much earlier than in the QM model (See Fig.~\ref{speedoiso2}).

In Fig.~\ref{speedoiso2}, we show the result for the speed of sound squared $c_s^2$ 
using the QM model for a wide range of values for $\mu_I/m_{\pi}$ (blue line).
The green band is the $\chi$PT result~\cite{qing} obtained as explained above.
The red band is the result of the lattice simulations of Ref.~\cite{abb24} with $m_{\pi}=139.6$ MeV and $f_{\pi}=92$ MeV. The band was constructed from 2000 bootstrap samples with \(\pm\) two standard deviations from the mean. The yellow band is the result of the
calculations in Ref.~\cite{minato}. The authors calculated the speed of sound, and trace anomaly using the  Cornwall-Jackiw-Tomboulis formalism~\cite{cjt}, including the gap in their two-loop calculations. Their band is obtained by varying the renormalization scale $\Lambda=X\mu$ with $X=1$ to $X=3$.
Our results show that $c_s^2$ increases sharply from the onset of pion condensation at $\mu_I=m_{\pi}$ with a peak at $\mu_I\approx 335$ MeV. After the peak, the speed of sound relaxes to the conformal limit (shown as the black dashed line).  As pointed out in Ref.~\cite{brandtqm}, the overall good agreement with lattice data over a wide range of isospin densities is remarkable. It is also in qualitative agreement with the perturbative calculation of~\cite{minato}.

\begin{figure}[htb!]
    \centering
    \includegraphics[width=\linewidth]{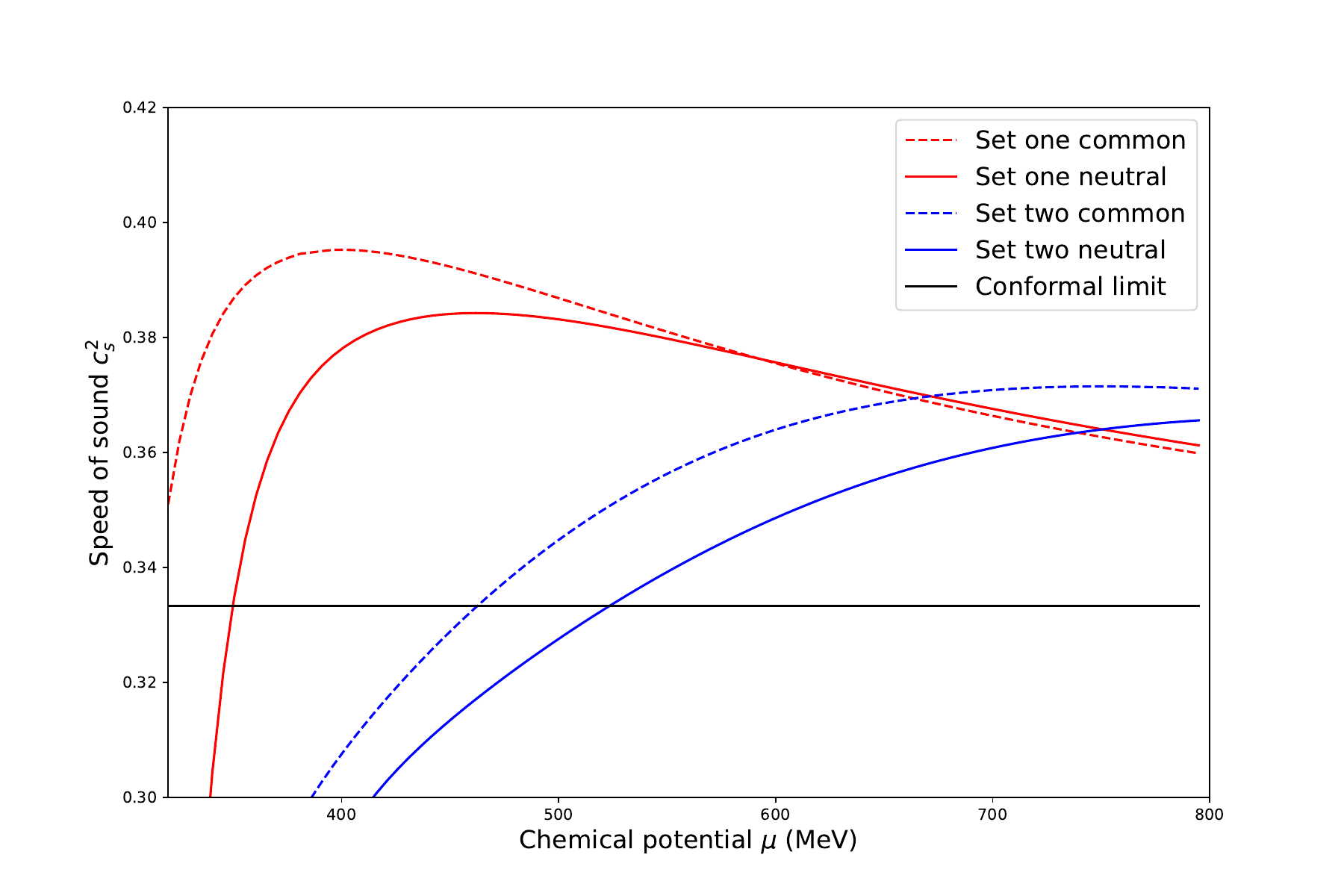}
    \caption{Speed of sound squared in the 2SC phase for parameter set 1 (red) and set 2 (blue) versus $\mu$. Dashed lines are without charge neutrality and solid lines are with charge neutrality. See main text for details.}
    \label{speedo}
\end{figure}

In Fig.~\ref{speedo}, we show the speed of sound squared $c_s^2$ as a function of the quark chemical potential. The red lines correspond to data set 1 and the blue lines refer to data set 2. Dashed lines are without electric and color charge neutrality, and solid lines are with. Imposing charge neutrality reduces the speed of sound for all values of $\mu$. Moreover, the peak of $c_s^2$ moves to higher values of $\mu$, and the peak is also lower. Thus, requiring charge neutrality softens the equation of state. 
We emphasize that this property and the relaxation to the conformal limit from above
are generic features of the QMD model. From Eq.~(\ref{speed2sc}) and the remarks
below it, it follows that $c_s$ with neutrality constraints imposed is larger than $c_s$ without charge neutrality, for sufficiently large values of $\mu$. This is the case for parameter set1,
as can be seen in Fig.~\ref{speedo}. The blue curves will cross each other for higher values 
of $\mu$.

Regarding the structure of $c_s$, we must be cautious, since our calculations are for two flavors. Extending our calculations to three flavors, we except to find a CFL phase for large $\mu$. If the transition from the 2SC phase to the CFL phase is first-order, the speed of sound will be discontinuous. Thus, we will not make a detailed comparison with other approaches~\cite{baymie}. That being said, a peak of $c_s^2$ has also been found in quarkyonic matter~\cite{quarkyonic}. On the other hand, the perturbative treatment of~\cite{minato} discussed above predicts that the conformal limit is approached from below.
Finally, the peak in the speed of sound is particularly interesting in view of recent neutron star observations. Using agnostic approaches, Refs.~\cite{perco,alti} find a peak, whereas Ref.~\cite{brandes} observes a plateau, as the baryon density increases.
It is possible to confront QMD model calculations with observation.
Combining mass-radius measurements of neutron stars with Bayesian inference~\cite{bayes1,bayes}, it is possible to find the probability distributions of its parameters.

Since we are interested in applying the QMD model to compact stars, we briefly discuss different possibilities for hybrid stars, i.e. stars with a core of deconfined quarks. In a compact star, a possible scenario is several phase transitions as we increase the baryon density. The first transition is from hadronic matter to normal quark matter.
The second transition is from normal quark matter to a color superconducting phase. Whether this transition is to the 2SC phase
or directly to the CFL depends on the details of the models (masses and couplings) and is an open question. One can ask the question whether the transition is sharp or is it via a mixed phase. A sharp transition takes place at a specific value for the
quark chemical potential at which the pressures of the two phases are equal. In each phase, neutrality constraints are imposed locally. 
The existence of a mixed phase requires a first-order transition between the two phases. In such a two-component system, neutrality
constraints are imposed globally~\cite{glendenning}. This means that the system is overall neutral with respect to the relevant conserved charges, but one component is positively charged and the other is negatively charged. Following Ref.~\cite{nonstrange}, we consider the possibility of a
mixed phase of normal quark matter and 2SC matter. Since normal quark matter is neutral with respect to the color charge $Q_8$, we also 
impose ${\partial\Omega_{0+1}\over\partial\mu_8}=0$ on the 2SC phase.
If the density in the core of the star is not too high, it is possible
that it contains a mixture of normal quark matter and 2SC matter, that is,
a non-strange hybrid star is realized. If the density in the core is sufficiently
high, one expects a transition to the CFL phase, so we need to extend the QMD model
to three flavors. Work in these directions is in progress~\cite{futurum}.

\section*{Acknowledgements}
The authors thank B. Brandt et al.~\cite{lattice}, R. Abbott et al.~\cite{abb24}, and K. Fukushima~\cite{minato} for permission to use their lattice data. J. O. A. also thanks B. Brandt, G. Endrodi, and L. von Smekal for useful discussions.\\ \\
{\it Note added:} In the final stages of this work, Ref.~\cite{brandtqm} appeared.
Their analysis of the pion-condensed phase and comparison with lattice simulations are similar to those of ours.



\bibliographystyle{apsrmp4-1}

\end{document}